\documentstyle[prl,aps,epsfig,psfig]{revtex}
\newcommand {\be} {\begin{equation}}
\newcommand {\ee} {\end{equation}}
\newcommand {\bea} {\begin{eqnarray}}
\newcommand {\eea} {\end{eqnarray}}
\newcommand{\bra}{\langle}
\newcommand{\ket}{\rangle}
\begin{document} 
\draft
\twocolumn[\hsize\textwidth\columnwidth\hsize\csname
@twocolumnfalse\endcsname
\widetext
\title{Electron-phonon interaction on bundled structures: 
static and transport properties}
\author{Ilaria Meccoli}
\address{Dipartimento di Fisica, Universit\'a di Parma,  
and Istituto Nazionale per la Fisica della Materia (INFM) - Unit\'a di Parma,
Parco Area delle Scienze 7a, I-43100 Parma, Italy}
\author{Massimo Capone}
\address{International School for Advanced Studies (SISSA), 
and Istituto Nazionale per la Fisica della Materia (INFM) \\ 
Unit\`a Trieste-SISSA, Via Beirut 2-4, I-34014 Trieste, Italy}
\date{\today}
\maketitle

\begin{abstract}

We study the small-polaron problem of a single electron interacting with
the lattice for the Holstein model in the adiabatic limit 
on a comb lattice, when the electron-phonon interaction 
acts only on the base sites. 
The ground state properties  can 
be easily deduced from the ones of a linear chain with an appropriate 
rescaling of the coupling constant. 
On the other hand, the dynamical properties, that involve the 
complete spectrum  of the system, present an "exotic" behavior. In the weak
coupling limit the Drude weight (zero-frequency conductivity) is enhanced 
with  respect to its free-case value, contrary to the linear chain case,
where for every finite value one has a suppression of the Drude peak.
More interestingly, the loss of coherent electron motion and the polaronic 
localization of the carrier occurs for different coupling values. Thus 
for intermediate coupling, a
novel phase appears with large kinetic energy and no coherent motion.

\end{abstract}
]
\narrowtext

\section{Introduction}
In the last years, a general interest has been growing around the study 
of physical properties of inhomogeneous discrete structures. 
The general purpose is
to characterize the behavior of systems like amorphous solids, 
glasses, polymers  and biological systems in general, 
where one can expect  the specific geometry  
to dramatically influence physical properties, ranging from transport 
and diffusive properties, to the thermal or electrical ones.
A very general and powerful formalism has been developed,
the so-called random walk problem, 
where the structure is explored by a classical walker, 
that  randomly jumps 
to an arbitrary  nearest-neighbor site at every time step. 
This approach can be thought as the discrete version 
of the classical Boltzmann equation, and the asymptotic behavior of the 
average distance of the walker from the starting site gives the diffusion 
law for the considered structure. On translationally invariant graphs, the 
diffusion law predicts that this distance scales with the time step with a 
power law depending on the euclidean dimension of the lattice. On more 
general graphs, the diffusion law is governed by a new parameter characterizing 
the structure, generally known as spectral dimension 
\cite{alexander,hattori,cassi&burioni}, 
that can be different from the euclidean dimension in which the graph is 
eventually embedded, giving rise to the so-called anomalous diffusion 
\cite{webman,havlin,sofia}. 
One of the main difficulties of this approach is 
that the explicit calculation of this important parameter may not be
trivial for general structures.

Nevertheless, there exists a wide class of graphs, called bundled 
structures, where explicit calculations can be performed. 
The random walk and the oscillation problems  have already been studied
on these structures\cite{sofia,oscill}, 
and an anomalous diffusion law for a classical walker moving
along the base has been discovered. 
Furthermore, the nearest-neighbor tight binding model has been explicitly
solved for electrons moving on these graphs, by means of an exact resummation 
of the perturbative expansion in the hopping parameter \cite{tb}. 
This perturbative approach becomes 
very complicated, and in a last instance useless, when the electrons 
experience any kind of interaction, both between them, and with
external fields, like e.g. phonons. 
This work is devoted to the first extension to the interacting case
of the problem of electrons moving on bundled structures.
We restrict ourselves to the case of a single electron which interacts
with local oscillators only on the base sites of a comb lattice. 
This is a first step towards the detection  of some quantum 
counterpart  of the anomalous diffusion laws.

The subject of electron-phonon (e-ph) interaction on unconventional
structures has been already addressed mainly within the framework of
the Su-Schrieffer-Heeger \cite{ssh} model, originally introduced
to describe the properties of poliacetylene chains \cite{librone}.
We will attack the problem of e-ph interaction on unconventional
structures from another point of view. Rather than a realistic 
description of actual compounds, we aim to provide a general picture
of the mutual effect of e-ph coupling and geometric complexity.
We will therefore focus on the Holstein model \cite{holstein}, which 
despite its simplicity, is expected to capture all the main physical
properties of more general interactions.
In section \ref{formalism} we sketch the basic formalism for bundled 
structures, and show how the problem can be mapped onto a base-only
one. In section \ref{e-ph} the small-polaron problem on the comb lattice
is discussed, with particular emphasis on the dynamical properties.
Section \ref{conclusions} is devoted to conclusions and future perspectives.

\section {General formalism}
\label{formalism}
A bundled structure can be built joining each point of a base graph 
${\cal B}$ with a copy of a fibre graph ${\cal F}$ in such a way that there is 
only one common site between ${\cal B}$ and ${\cal F}$. 
As previously explained, we shall limit our study to the 
simplified case in which the particles experience interaction 
only on the base sites.
Within this hypothesis, the generic hamiltonian for a bundled 
structure assumes a block form:
\be
\left(
\begin{array}{cc}
H_B~~~ H_{BF}\\
H_{FB}~~~H_F
\end{array}
\right)
\label{ham}
\ee
$H_B$ and $H_F$ act on the base and the fibres sites respectively, 
while $H_{BF}=H^{T}_{FB}$ 
are rectangular matrices containing the hopping terms joining 
the fibres to the base. Consequently, a generic state can be written as
\be
|\phi\ket =\left(
\begin{array}{ll}
|\psi \ket \\
|\eta \ket 
\end{array}
\label{ket}
\right)
\ee
with $|\psi\ket$ belonging to the base and $|\eta\ket$ to the fibres. 
Since this separation of degrees of freedom is completely formal, 
we must ensure that it does not introduce  unphysical states in the 
Hilbert space. So we impose a normalization constraint: 
$\bra \psi |\psi \ket + \bra \eta |\eta \ket= \bra \phi|\phi \ket=1$.
Let us write the eigenvalues equation for this structure:
\bea
&&H_B |\psi_\epsilon \ket +H_{BF}|\eta_\epsilon \ket= 
\epsilon |\psi_\epsilon \ket
\nonumber\\
&&H_F |\eta_\epsilon \ket +H_{FB}|\psi_\epsilon \ket= 
\epsilon |\psi_\epsilon \ket
\label{shr1}
\eea
Because $H_F$ does not contain any interaction term, one can get rid of 
$|\eta_\epsilon \ket$ from (\ref{shr1}), reducing to a ``base-only'' problem:
\begin{eqnarray}
\label{pot&norm}
( \epsilon -H_B - H_{BF}(\epsilon -H_F)^{-1}H_{FB})
|\psi_\epsilon\ket=0
\nonumber\\
\langle \psi_\epsilon\vert [1-H_{BF}
(\epsilon -H_F)^{-2}H_{FB} ]\vert\psi_\epsilon \rangle = 1
\end{eqnarray}
Explicitly using the simple form of the rectangular matrix $H_{BF}$ 
(reflecting
the peculiar geometry of the bundled structure), it can be 
shown that the composite operators in (\ref{pot&norm}) have only diagonal 
terms simply expressed in term of the fibre-only Green's functions:
\bea
&&\left[H_{BF}(\epsilon -H_F)^{-1}H_{FB}\right]_{ij}=
\delta_{ij}n_f t^2 F_{00}(\epsilon)
\label{pot}\\
&&\left[H_{BF}(\epsilon -H_F)^{-2}H_{FB}\right]_{ij}=
\delta_{ij}n_f t^2 \sum_m F_{0m}(\epsilon)F_{m0}(\epsilon)
\label{norm}
\eea
Here $n_f$ is the number of independent fibres joined to a single site of the 
base; $t$ is the hopping constant relative to the link joining the 
fibre and the base; $F_{lm}(\epsilon)= [\epsilon -H_F]^{-1}_{lm}$ is 
the fibre Green's function , and $i=0$ indicates the first site of the 
fibre when coming from the base. So we have reduced ourselves to an 
effective problem for the only base sites:
\bea
&&\left( \epsilon -H_B - n_f t^2 F_{00}(\epsilon)\right)|\psi_\epsilon\ket=0
\label{eff.schr}\\
&&\bra \psi_\epsilon |\psi_\epsilon \ket=
\frac{1}{1+n_f t^2\sum_m F_{0m}(\epsilon)F_{m0}(\epsilon)}\equiv n(\epsilon).
\label{eff.norm}
\eea
We can rewrite Eq.(\ref{eff.schr}) as
\be
\label{effe}
\left( f(\epsilon) -H_B \right)|\psi_\epsilon\ket=0,
\ee
defining $f(\epsilon) = \epsilon -  n_f t^2 F_{00}(\epsilon)$.
Two effects keep track of the  presence of the fibres.
A time-dependent potential 
$V(\tau)=n_f t^2\int d \tau' \psi^*(\tau)F_{00}(\tau - \tau')\psi(\tau ')$, 
of purely geometrical origin, acts on
 the base, accounting for  the possibility for the particle
 to explore the fibres when moving on the base. On the other hand, 
the normalization of the eigenvectors $n(\epsilon)$ depends 
self-consistently on 
the corresponding eigenvalue. This fact will have a dramatic 
consequence in presence of interaction, and the
 Schr\"odinger equation contains a nonlinear term. Both these 
aspects will be analyzed in the subsequent section.

Let us note that the decomposition (\ref{ham}, \ref{ket}) holds 
under the very general hypothesis that the fibres and the base 
are connected only by an hopping term. If $H_F$ 
contains instead an interaction term, 
we generally not know the explicit form 
of the Green's functions $F_{ij}(\epsilon)$ and consequently the 
expressions (\ref{pot}), (\ref{norm}) can be not be computed exactly.

\section{Electron-phonon interaction: the comb lattice case}
\label{e-ph}
In this section  we turn to the problem of electron-phonon interaction
on bundled structures. We will consider the simplest model for
e-ph interaction, namely the Holstein molecular crystal model \cite{holstein}. 
\begin{eqnarray}
\label{hamiltonian}
H &=& -t\sum_{\langle i,j \rangle ,\sigma} 
(c_{i\sigma}^{\dagger} c_{j\sigma} + h.c) +\nonumber\\ 
&+&\tilde{g}\sum_i n_i (a_i + a_i^{\dagger}) + 
\omega_0\sum_i a_i^{\dagger} a_i.
\end{eqnarray}
In this model tight-binding electrons interact with local dispersionless
oscillators of frequency $\omega_0$, $\tilde{g}$ is an
e-ph coupling between the displacement $a_i + a_i^{\dagger}$ of the 
oscillator and the local electronic density
$n_i = \sum_{\sigma} c_{i\sigma}^{\dagger} c_{i\sigma}$.
On really general grounds, the model is expected to display a polaronic 
behavior for large el-ph coupling. A polaron is a bound state of electrons 
and phonons in which the electron moves carrying with itself a phonon cloud 
that strongly diminishes its mobility, eventually giving rise to localization 
\cite{holstein,emin}.
Since, at least for $\tilde g = 0$, the electron motion is completely free,
some kind of transition or crossover from a weak-coupling
free-carrier state, and a strong-coupling polaronic state
must occur at some coupling.
Despite the simplicity of the model,
even the case of a single particle interacting with the set of
oscillators is a non-trivial many-body problem that can be solved 
only in particular limits. 
We mention the solution obtained in 
Ref.\cite{ciuk} within the Dynamical Mean Field, which is exact
in the limit of infinite spatial dimensions, and the theorem
by Gerlach and L\"owen that claims that no phase
transition occurs for finite $\omega_0$ \cite{gerlach_and_lowen}.
In general cases one has to resort to
approximate techniques or numerical studies\cite{csg,ccg,stephan}.

The nature of the transition between a free carrier (for $\tilde{g}=0$)
and a polaron depends on the value of $\omega_0/t$ and spatial dimensionality
\cite{ccg}.
In fact, 
even for a single particle we have two independent parameters 
in the Hamiltonian  (\ref{hamiltonian}), which determine different 
physical regimes. In particular, while $\tilde{g}$ is a measure of 
the strength of the e-ph interaction, $\gamma = \omega_0 /t$ is 
a measure of the adiabaticity of the system, i.e. it controls the 
relative value of the typical 
phononic and electronic energy scales. It has been shown that in the 
adiabatic regime $\gamma \ll 1$, the condition that  rules the polaronic 
crossover is $\lambda = {\tilde g}^2/(2\omega_0 t d) > 1$, where $d$ is
the dimensionality.
This condition is simply
understood as an energy convenience condition as soon as we notice
that the polaronic energy is $E_{pol} = -\tilde g^2/\omega_0$, and
the free carrier energy is $E_{free} = -2td$. 
Notice that $E_{free}$ contains the total kinetic energy\cite{csg,ccg}.

We will restrict ourselves to the extreme adiabatic limit $\omega_0/t 
= 0$, in which the lattice degrees of freedom become classical.
In this limit, the Hamiltonian can be recast in the form
\begin{equation}
\label{hamadi}
H = -t\sum_{\langle i,j \rangle ,\sigma}
(c_{i\sigma}^{\dagger} c_{j\sigma} + h.c) +
g\sum_i n_i x_i+{1\over 2}k\sum_i x_i^2.
\end{equation}
Here $g=\sqrt{2k/\omega_0}\tilde{g}$, $k$ is the elastic constant, 
and $x_i$ are the classical variables
associated to the local displacements of the oscillators.
It can be shown that the relevant adimensional coupling
the adiabatic limit is $\lambda = g^2/4kt$.

Kabanov and Mashtakov\cite{km} have solved the model for 
a single particle for 1-2-3 dimensional systems showing that in 
the one-dimensional case the system has a localized bound state 
for arbitrarily small e-ph coupling, evidencing only a crossover 
between a weak-coupling large
polaron extended on many lattice sites, and a strong-coupling small 
polaron localized on a single site. In two and three spatial dimensions
a first-order transition (level crossing) occurs between a delocalized state
and a small polaron one.
We will generalize the approach of reference \cite{km} to the
case of a comb lattice, a specific bundled structure, in which 
the base is a linear chain, and two semi-infinite chains are linked
to each site of the chain (fibres).
We will also consider only the case in which the e-ph 
interaction is limited to the base. 
This lattice can be viewed as a good model for describing the geometry of 
many branched polymers \cite{polymers}, and, on the other side, it is simple 
enough to obtain explicit results.

In order to reduce ourselves to a base-only problem, we can follow the 
procedure outlined in the previous section. 
Attention must be payed to the fact that $\epsilon$ indicates only the 
contribution of the electronic degrees of freedom to the full 
eigenvalue and the energy is given by $\epsilon + \frac{1}{2}k\sum_i x_i^2$.
The Schr\"odinger equation for the base degrees of freedom can be 
obtained from the one in Ref. \cite{km} in two steps.
First of all, the electron eigenvalue $\epsilon$ must be substituted with  
$f(\epsilon)$, defined in Eq. (\ref{effe}), a quantity that keeps track of
the  effects of the fibres on the electron dynamics.
The second step amounts to re-normalize the wavefunction 
according to Eq.  (\ref{eff.norm}), i.e. 
$\psi_i \rightarrow \psi_i'/\sqrt{n(\epsilon)}$\cite{footnote}. 
We then obtain, following Kabanov and Mashtakov \cite{km},
\be
\left(f(\epsilon)+{2\lambda t} n(\epsilon)|\psi'_i|^2 \right)\psi'_i=
t \sum_l \psi'_{i+l}
\label{km_eff}
\ee
It is now clear that the condition (\ref{eff.norm}) introduces 
 an effective rescaling of the coupling constant 
$\lambda = g^2/4kt$. For $n(\epsilon)\leq 1$, the interaction 
term is less effective with respect to the base-only problem. 
Furthermore, every energy level can be mapped onto one of the base-only 
problem,  with a coupling changing self-consistently with the energy value.

As a consequence, the physical quantities involving expectation values on a 
single eigenfunction will have similar behaviors for the bundled structure 
and the base-only cases. On the other hand, we expect different 
properties for quantities which involve matrix elements between different
eigenfunctions, like e.g. spectral properties.

In what follows, we choose the explicit case of the comb lattice, 
comparing our results with the ones of the one-dimensional case. 
The fibre is a semi-infinite linear chain and the 
expressions (\ref{pot}), (\ref{norm}) take the explicit form:
\bea
&&F_{00}(\epsilon)=\frac{1}{2t^2}\left( \epsilon -sgn(\epsilon)\sqrt{\epsilon^2 -4t^2}\right)
\label{f00}\\
&&n(\epsilon)=\frac{\sqrt{\epsilon^2 -4t^2}}{|\epsilon|}
\label{epsi}
\eea 
Looking at eq. (\ref{km_eff}), we expect that, due to the effective 
reduction of the coupling constant, the crossover from a large to a 
small polaron, typical of the one dimensional case, will occur for 
a larger value of $\lambda$. Moreover, the ground state properties 
of the one-dimensional case can  be recovered from the ones of the 
comb lattice through a redefinition of the coupling
and a rescaling of the wavefunction:
\bea 
&&\lambda \rightarrow \lambda n(\epsilon_{\lambda}) \nonumber \\ 
&& \psi_i \rightarrow \psi_i/\sqrt{n(\epsilon_{\lambda})}\label{scaling}\\
&& x_i \rightarrow x_i/\sqrt{n(\epsilon_{\lambda})}\nonumber.
\eea
Here we emphasize the dependence on $\lambda$ of the eigenvalues
$\epsilon_{\lambda}$.

\begin{figure}
\centerline{\epsfig{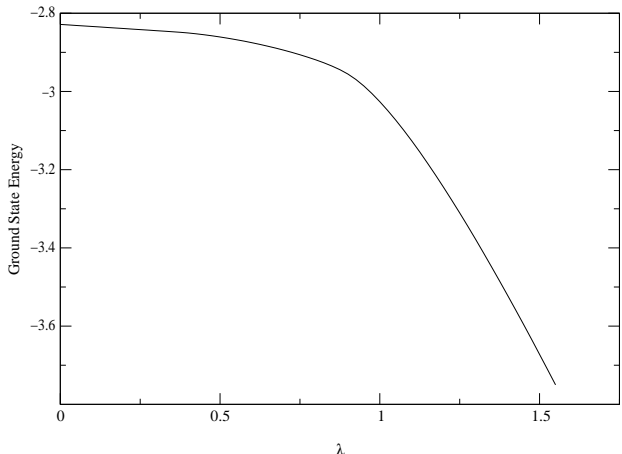}}
\caption{Ground state energy for a $20\times 20$ comb lattice 
(in units of $t$), as a function of the coupling constant $\lambda$.}
\label{energia}
\end{figure}

We have also investigated if the rescaling of the eigenvalue equation 
could induce a level crossing in the ground state at some value of the 
coupling, thus modifying the  nature of the crossover with respect to 
the $1d$ case. 
The first derivative of $\epsilon(\lambda)$ turns out to be  proportional 
to the same  quantity of the base-only case, through a continuous function 
of $\epsilon$,  built up by $F_{00}(\epsilon)$ and its derivative. 
In this way we conclude that the properties of the ground state for a 
comb lattice, or generally for a bundled structure whose fibres are 
semi-infinite linear chains, are the same of the ones of the base-only case.

We performed numerical solution of the problem for finite, but quite large 
lattices, up to 20$\times$ 20 sites.
Henceforth, energies will be expressed in units of the hopping amplitude $t$,
and lengths in units of the lattice constant $a$.
Our numerical data perfectly confirm the previous considerations. 
In figure \ref{energia} we plot the ground state energy as a function 
of the coupling constant $\lambda$. It presents a continuous crossover
between a large polaron state whose energy is only slightly modified 
with respect to  the free value $\epsilon_0 =-2\sqrt{2} t$ to a strong 
coupling small polaron state, in which $\epsilon_0  = -2t\sqrt{2}\lambda$. 
This behavior is completely analogous to the one-dimensional 
system\cite{km}.

\begin{figure}
\centerline{\epsfig{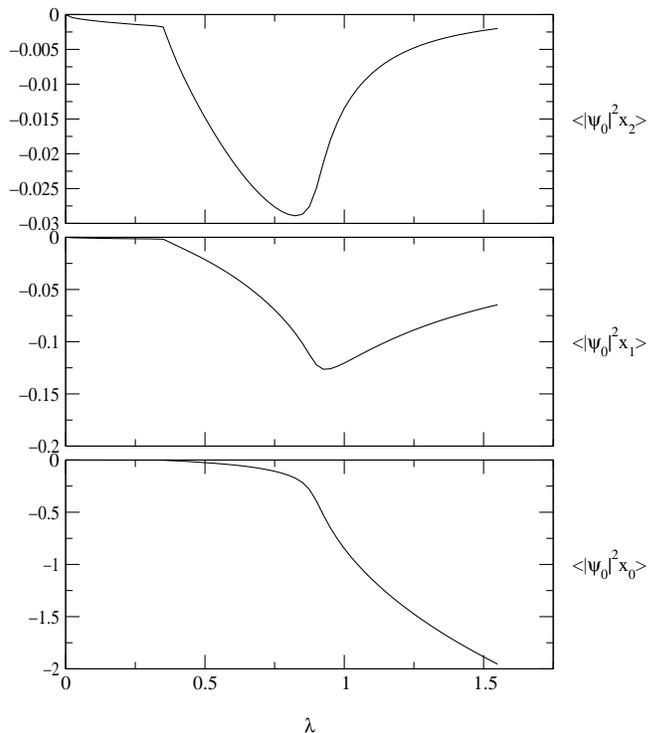}}
\caption{Density-displacement correlation functions for a $20\times 20$
comb lattice for different 
distances of the sites from the density peak. The change of the slope 
of the local correlation function (bottom panel) is 
the signal of the polaronic crossover, and permits to extract the critical 
value $\lambda_c=0.925$.}
\label{corr}
\end{figure}

In order to estimate the crossover value $\lambda_c$, 
we plot the density-displacement correlation functions 
$\bra |\psi_0|^2 x_i\ket$ for $i=0,~1,~2$ (fig. \ref{corr}), 
where $i=0$ indicates the site where the electron localizes \cite{csg}.
This quantity is a direct probe of the polaronic behavior,
measuring the degree of correlation between the electron and the
lattice deformations.
All the non-local correlation functions decay rather sharply
above a given value of $\lambda$. Roughly at the same coupling value,
the local correlation function abruptly changes the slope, 
signaling a crossover from a weak-coupling to a strong-coupling
regime. This value can be associated
to the crossover from a large polaron extended on several lattice
site to a single-site small polaron.

We can therefore estimate the crossover coupling $\lambda_c$ with
the value where the first derivative of the local 
correlation function is maximum, or equivalently with the value
for which the nearest-neighbours correlation function starts to
decrease.  
Using the first criterion we obtain an estimate $\lambda_c=0.925$, 
which has to be compared with 
the one-dimensional value $\lambda_c =0.75$. 

Notice that our comb lattice can be viewed as a first step beyond 
one-dimensionality. The crossover coupling for $d=2$ is $\lambda \simeq 1$. 
One should anyway keep in mind that we are not considering el-ph 
interaction on the fibres.

\begin{figure}
\centerline{\epsfig{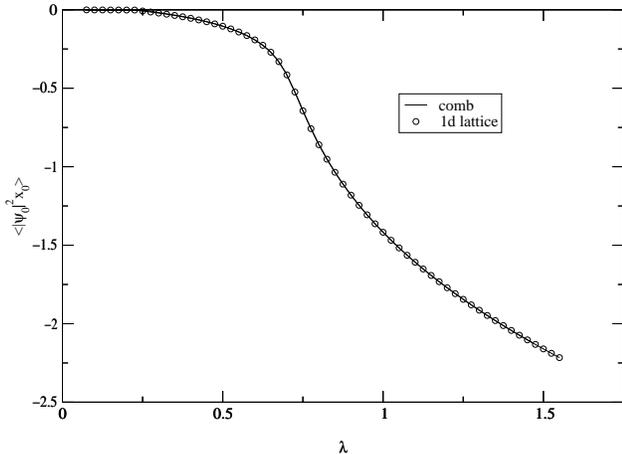}}
\caption{The local density displacement correlation function for the 
linear chain is derived from the data by the comb lattice case with 
the scaling explained in the text (See Eq. (\protect\ref{scaling})).} 
\label{scaling_corr}
\end{figure}

In Fig. \ref{scaling_corr} we show how the $1d$ quantities can be 
recovered with the scaling relations (\ref{scaling}):
Once the local density-displacement correlation function
$\langle\vert\psi_0\vert^2 x_0\rangle$
and $\lambda$ are scaled respectively by $n^{3/2}(\epsilon_{\lambda})$ and 
$n(\epsilon_{\lambda})$, the curve lies on top of the one for the linear chain.

As expected no qualitative difference with respect to the one-dimensional
case has been found for the groundstate
properties.

\subsection{Transport properties}

The study of the transport properties in the polaronic problem has been 
widely studied \cite{akr,csg}, since it directly characterizes the onset of 
localization.

One of the most striking features of bundled structures is an anomalous 
diffusion for a classical walker moving along the base \cite{sofia}.
We expect quantum counterpart for this property, namely some peculiar 
behavior for the conductivity. Note that this expectation is not in 
contradiction with the general properties of the eigenstates described 
in the previous section, since the optical response is expressed in 
terms of matrix elements between {\it different} eigenstates. In fact 
the mapping (\ref{scaling}) involves different normalizations, and 
therefore different effective coupling constants, for every energy level.

\begin{figure}
\centerline{\epsfig{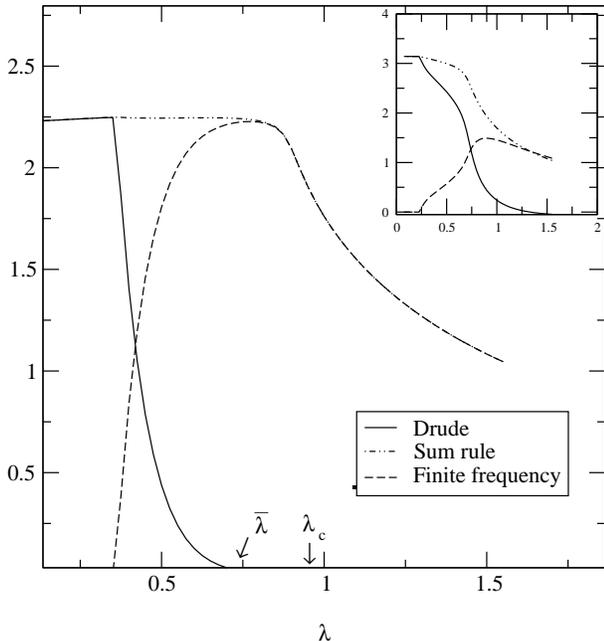}}
\caption{Contributions to the DC conductivity along the backbone,
compared with the 1d case. In both cases we show the Drude weight $D$, 
the total sumrule for the optical conductivity 
(\protect\ref{sumrule}), and the finite frequency optical weight.}
\label{figweights}
\end{figure}

The optical conductivity is given by:
\be
\sigma(\omega)= D\delta(\omega)+
\sum_{n\ne 0}\frac{|\langle \psi_n|J_x|\psi_0 \rangle|^2}{\epsilon_n -
\epsilon_0}
\label{conductivity}
\ee
where the coefficient of the zero-frequency delta function 
contribution is the so called Drude weight, which explicitly 
characterizes the transport properties: a vanishing $D$ is the signature 
of an insulating state \cite{kohn}, whereas a metal has a finite value of 
such a quantity. The Drude peak can be evaluated by means of the so called 
$f$-sumrule \cite{mald}:
\be
\int_{0}^{\infty}\sigma(\omega) d\omega =-\frac {\pi e^2}{2}
\langle H_{kin}\rangle
\label{sumrule}
\ee
and involves the calculation of the kinetic energy and of the integral
over all finite frequencies of the optical conductivity.
Quite naturally we only consider the conductivity along the backbone, 
in which anomalous behavior can be expected.
We compute the Drude weight given by (\ref{conductivity}), 
the total weight of the optical excitations, 
related to the kinetic energy by Eq. (\ref{sumrule}),
and the integral over finite (non-zero) frequencies.
Both the kinetic energy and the finite frequency optical conductivity
 show an anomalous behavior with respect to the 
one dimensional case as it is transparent from a comparison
between the data in Fig. \ref{figweights} and the ones for
the one-dimensional case reported in the inset.

For small values of the coupling constant the absolute value of the 
kinetic energy presents a small but visible enhancement with respect to 
its free value. This is confirmed by a perturbative calculation for 
$\lambda \ll 1$. In this limit, we can introduce a continuous 
approximation of eq. (\ref{km_eff}) \cite{km,akr}:
\be
\left(\alpha+2t+{2\lambda t} |\psi(x)|^2\right)=
-a^2 t \frac{d^2 \psi(x)}{d~x^2}
\ee
 which has an  exact solution of the form:
\be
\psi(x)=\sqrt{\frac{\lambda}{2a}}
\frac{n(\epsilon_{\lambda})}{\cosh(\lambda n(\epsilon_{\lambda})x/a)}
\\
\ee

In the one-dimensional case $n(\epsilon_{\lambda})=1$ and $\alpha$ just 
represents the electronic contribution to the ground state energy, 
while for the comb lattice they are given by Eqs. (\ref{f00}), 
and (\ref{epsi}). 
Thus the averaged kinetic energy  is $E_{KIN}=-2t(1-\lambda^2 t/6a^2)$ 
for the $1d$ case, while 
$E_{KIN}\simeq-\sqrt{2}t(1+\lambda^2 t/8a^2-\lambda^4/128a^4)$ 
on the comb lattice. 
This result can be easily understood by inspection of  the particular 
geometry of the lattice. 
In the free case, the ground state wavefunction has a constant 
value on the backbone sites with exponentially decreasing tails on the 
teeth. Switching on the interaction, the backbone sites become energetically 
favorable and  the particle is therefore recalled  on the backbone, 
increasing  the charge density on the substructure. 
and consequently of the kinetic energy. 
Further increasing $\lambda$ the localization tendency which leads
to small polaron formation becomes effective. In this regime
the usual behavior is recovered, the kinetic energy decreases 
its absolute value, while the e-ph interaction energy is 
substantially decreased. 

For small values of $\lambda$ the finite frequency contribution 
remains zero up to a given value of $\lambda$, 
and continuously acquires a finite value at some value of $\lambda$
for both the structures. This can be attributed to finite size effects: 
a finite number of sites cuts off the available states 
with proper symmetry in (\ref{conductivity}).

A striking feature emerges from Fig. \ref{figweights}. 
The coupling value for which the Drude weight vanishes 
($\bar\lambda \simeq 0.72$) is sizeably
smaller than the crossover coupling extracted from the correlation
functions ($\lambda_c \simeq 0.925$), 
which in turn coincides with the kink in the kinetic energy curve. 
This is never the case in usual one-dimensional and
two-dimensional lattices, where the loss of coherence and the 
small polaron formation essentially coincide, as can be seen in the inset, 
and as it is valid also for finite phonon frequencies.

However, our results are not in contradiction with general criteria for 
small polaron formation \cite{csg} but they confirm a crucial physical point.
In fact the condition 
$\lambda > 1$ implies that the electronic energy to 
be overcome by the  el-ph interaction energy for small polaron 
formation is not the coherent one but the total kinetic energy. These 
quantities coincide for a single particle on a conventional lattice, 
but they may differ for interacting systems or unordered structures.

For $\bar \lambda < \lambda < \lambda_c$ the system is in a state 
in which the coherent transport is strongly 
suppressed while the kinetic energy is still large, 
substantially equal to its non interacting value. 
We attribute this contribution to local
incoherent electron hopping between the teeth sites.

This system represents, to our knowledge, the first one in which  
coherent motion is strongly inhibited, but there is a sizeable range of 
parameters in which no polaronic effects are found. This phenomenon represents 
the quantum heir of anomalous diffusion. 
It should be noted that, as the el-ph coupling is extended  to the whole 
lattice this effect is expected to be emphasized.

\section{Conclusions and perspectives}
\label{conclusions}
The static and dynamical properties of an electron interacting with
local classical oscillators on a comb lattice have been extensively
analysed. 
Even though our work represents only the first attempt to tackle 
the problem of interacting fermions on unconventional structures,
interesting and peculiar features have been discovered.
On really general grounds, we have shown that the static properties
(averages on a single state) of a general bundled structure
can be easily obtained from the ones of the base-only problem 
by means of a rescaling of the parameters.
As an example, we have studied the small polaron crossover on a comb
lattice, comparing it with the one-dimensional case, explicitly
confirming this property.
Nevertheless, dynamical properties cannot be so simply recovered.
A numerical study of the optical response on the comb lattice has
highlighted two anomalous properties.
Contrarily to the one-dimensional case, for small e-ph couplings, the
DC conductivity (measured by the Drude weight) is increased with respect 
to the bare non-interacting value.
By increasing the coupling a rather surprising feature has been found,
namely the coherent motion is strongly suppressed while a sizeable
incoherent kinetic energy is still present. 
The loss of coherence is therefore not due to the self-trapping
associated with the polaron crossover (that occurs for larger 
e-ph coupling, when also the incoherent kinetic energy is suppressed),
but to the drastic effects of a unconventional geometry on the 
electronic properties.
The occurrence of such an effect in classical systems has been 
demonstrated  and widely discussed \cite{sofia} within the framework
of the random walk.
We note that our approach is not completely the analogous of the random 
walk problem: the absence of e-ph interaction on the teeth sites 
implies that, in the classical analogous, the motion of the walker on these 
sites could be deterministic. The quantum counterpart of 
anomalous diffusion driven by the 
geometry would be achieved only if the e-ph interaction
was extended to the whole lattice.

\section{Acknowledgments}
We wish to thank G. Burgio, R. Burioni and D. Cassi for helpful discussions,
S. Caprara for discussions and a careful reading of the manuscript.

\bibliographystyle{prsty}
\bibliography{pre}

\begin{thebibliography}{1}
\bibitem{footnote}It is clear that the same considerations can be 
applicable every time the considered interaction depends on the 
electron density alone.
\end{thebibliography}

\end{document}